\documentclass[draft]{agujournal2019}
\usepackage{url} %this package should fix any errors with URLs in refs.
\usepackage{lineno}
\usepackage[inline]{trackchanges} %for better track changes. finalnew option will compile document with changes incorporated.
\usepackage{soul}
\usepackage{subcaption}
\usepackage{graphicx}
\usepackage{amsmath}
\nolinenumbers
\draftfalse

\journalname{JGR: Space Physics}

\begin{document}

%% ------------------------------------------------------------------------ %%
%  Title
%
% (A title should be specific, informative, and brief. Use
% abbreviations only if they are defined in the abstract. Titles that
% start with general keywords then specific terms are optimized in
% searches)
%
%% ------------------------------------------------------------------------ %%

\title{Numerical Modelling and GNSS Observations of Ionospheric Depletions due to a Small-Lift Launch Vehicle}

%% ------------------------------------------------------------------------ %%
%
%  AUTHORS AND AFFILIATIONS
%
%% ------------------------------------------------------------------------ %%

\authors{G. W. Bowden \affil{1}, M. Brown \affil{1}}

\affiliation{1}{School of Engineering and Information Technology, University of New South Wales Canberra, Campbell, ACT, Australia}

\correspondingauthor{George Bowden}{g.bowden@adfa.edu.au}

%% Keypoints, final entry on title page.

%  List up to three key points (at least one is required)
%  Key Points summarize the main points and conclusions of the article
%  Each must be 140 characters or fewer with no special characters or punctuation and must be complete sentences

\begin{keypoints}
\item Ionospheric effects of the Rocket Lab Electron launch of 22 March, 2021 were investigated
\item Numerical modelling using GITM and navigation satellite signal observations showed comparable ionospheric depletions
\end{keypoints}

%% ------------------------------------------------------------------------ %%
%
%  ABSTRACT and PLAIN LANGUAGE SUMMARY
%
% A good Abstract will begin with a short description of the problem
% being addressed, briefly describe the new data or analyses, then
% briefly states the main conclusion(s) and how they are supported and
% uncertainties.

% The Plain Language Summary should be written for a broad audience,
% including journalists and the science-interested public, that will not have 
% a background in your field.
%
% A Plain Language Summary is required in GRL, JGR: Planets, JGR: Biogeosciences,
% JGR: Oceans, G-Cubed, Reviews of Geophysics, and JAMES.
% see http://sharingscience.agu.org/creating-plain-language-summary/)
%
%% ------------------------------------------------------------------------ %%

%% \begin{abstract} starts the second page

\begin{abstract}
Space launches produce ionospheric disturbances which can be observed through measurements such as Global Navigation Satellite System signal delays. Here we report observations and numerical simulations of the ionospheric depletion due to a Small-Lift Launch Vehicle. The case examined was the launch of a Rocket Lab Electron at 22:30 UTC on March 22, 2021. Despite the very small launch vehicle, ground stations in the Chatham Islands measured decreases in line-of-sight total electron content for navigation satellite signals following the launch. General Circulation Model results indicated ionospheric depletions which were comparable with these measurements. Line-of-sight measurements showed a maximum decrease of $2.7$~TECU in vertical total electron content, compared with a simulated decrease of $2.6$~TECU. Advection of the exhaust plume due to its initial velocity and subsequent effects of neutral winds are identified as some remaining challenges for this form of modelling.
\end{abstract}

\section*{Plain Language Summary}
The ionosphere is a region of the upper atmosphere containing a mixture of ions and electrons. Chemicals released when rockets pass through this region allow ions and electrons to recombine, reducing their concentrations. These reductions were measured for the launch of a Rocket Lab Electron rocket. The measurements were made based on the difference in time signals at different frequencies took to pass through the ionosphere, which indicated its concentration. Computer simulations were carried out for this case, which showed a similar decrease in the concentration of the ionosphere following launch to the the measurements. Some potential improvements to the models are suggested based on this work.

%% ------------------------------------------------------------------------ %%
%
%  TEXT
%
%% ------------------------------------------------------------------------ %%

\section{Introduction}
During space launch, rockets release exhaust gasses which modify the ionosphere. The exhaust includes heteronuclear molecules such as $\text{CO}_2$ and $\text{H}_2\text{O}$, which undergo rapid charge exchange reactions with $\text{O}^+$ in the $\text{F}_2$-region to form molecular ions \cite{mendillo1981}. These ions readily undergo neutralisation reactions, reducing ionospheric electron densities. Ionospheric depletions following rocket launches have been observed using a variety of methods, including ionosonde soundings \cite{booker1961}, Faraday rotation of satellite signals \cite{mendillo1975}, air glow images \cite{mendillo1982}, incoherent scatter radar measurements \cite{wand1984}, Global Navigation Satellite System (GNSS) signal delays \cite{furuya2008}, and satellite Langmuir probe measurements \cite{park2016}. These depletions can provide a means of investigating ionospheric physics or detecting and characterising launches.

% https://www.rocketlabusa.com/launch/electron/
In this study, we investigate ionospheric depletions due to the launch of an Rocket Lab Electron launch vehicle. This three-stage partially recoverable Small-Lift Launch Vehicle (SLLV) has a wet mass of $13,000$~kg and carries payloads of up to $300$~kg to low-Earth orbit (LEO) \cite{rocketlab2022}. Both stages use an RP-1/LOX propellant mixture. Observations of ionospheric depletions were first made for the launch of the Vanguard SLV-4 rocket, a comparably sized SLLV \cite{booker1961}. However, ionospheric effects of such small launch vehicles have not previously been simulated numerically or observed in GNSS data.

The launch (mission name ``They Go Up So Fast") occurred at 22:30 UTC on March 22, 2021, corresponding to 11:15 Chatham Standard Time (CHAST). Rocket Lab Launch Complex 1 was the launch site, located at $39.262^\circ$S, $177.865^\circ$E (on the M\={a}hia Peninsula, New Zealand). The launch vehicle carried 7 small satellites to target altitudes between $450$~km-$550$~km and $45.0^{\circ}$ inclination. Section~\ref{sec:numerical_modelling} and \ref{sec:gnss_observations} respectively outline the numerical ionosphere-thermosphere system modelling and GNSS observation methods applied to the ``They Go Up So Fast'' launch case.  Results of these investigations are detailed in Section~\ref{sec:results} and further discussed in Section~\ref{sec:discussion}.

\section{Numerical Modelling}
\label{sec:numerical_modelling}
A numerical simulation of the ionospheric depletion due to the rocket launch was performed using the Global Ionosphere Thermosphere Model (GITM) \cite{ridley2006} adopting the methods outlined by \citeA{bowden2020}. This simulation covered an 8-hour period from 22:30 UTC on March 22 to 06:30 UTC on March 23. The chosen domain spanned $50^\circ$S to $30^\circ$S in latitude, $170^\circ$E to $155^\circ$W in longitude, and $100$~km to $536.67$~km in altitude. This domain was covered by a regular $108 \times 144$ grid horizontally and $50$ grid points vertically. These altitudes corresponded to the thermosphere (in the neutral atmosphere) and E- and F-regions (in the ionosphere). The simulation was run both with and without rocket exhaust gasses being added. The electon density integrated along the line-of-sight, or Slant Total Electron Count (STEC), was computed for GNSS satellites based on azimuth and elevation data from the source described in Section~\ref{sec:gnss_observations}. Output was taken at $300$~s intervals.

The rocket trajectory was estimated based on the target orbit, altitude, and speed data. Rocket Lab provide speed and altitude data in their launch video \cite{youtube2021}. These data indicated that passage through the simulation domain coincided with the second stage firing. The total mass flow rate during the second stage firing was estimated based on a nominal thrust of $T = 25.8$~kN and specific impulse of $I_{sp} = 343$~s using $\dot{m} = T / \left ( I_{sp} g_0 \right )$ \cite{rocketlab2022}. This was divided between $\textup{H}_2\textup{O}$ and $\textup{CO}_2$ assuming complete combustion. The rocket chemical source was added at points along the trajectory for $520$~s following launch (approximately corresponding to second stage cut-off at approximately $320$~km altitude). These were added following $300$~s delay during which diffusion was modelled analytically, avoiding excessive concentration gradients which cause problems for the GITM numerical solver \cite{bowden2020}.

\section{GNSS Observations}
\label{sec:gnss_observations}
\begin{figure}
\includegraphics[width=120mm]{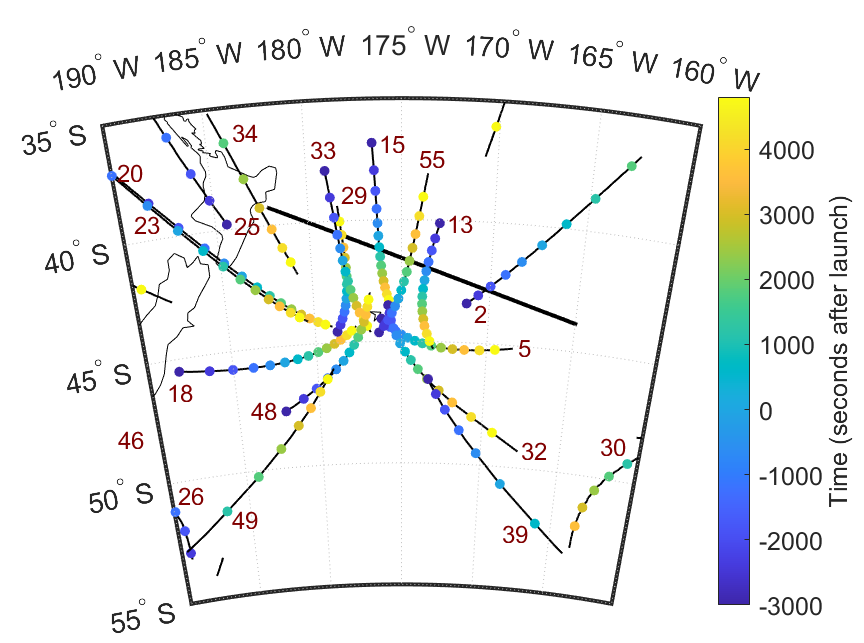}
\centering
\caption{Pierce points for TEC measurements from the Chatham Islands GNSS ground station (CHTI) between $21$:$40$ and $24$:$00$~UTC. Markers are plotted at $10$~minute intervals The estimated rocket ground track while the first and second stages were firing is indicated by the thick black line.}
\label{fig:GNSS_Ground_Tracks_CHTI}
\end{figure}

Changes in Total Electron Content (TEC) following the launch were measured from the Chatham Islands reference station (CHTI), located at $43.735^\circ$S, $176.617^\circ$E, $75.764$~m altitude. Slant TEC (STEC) data for the CHTI station were obtained from the Madrigal CEDAR database. These were available for both GPS (here numbered 1 to 31) and GLONASS (numbered 32 to 55) satellites. Data also included estimated pierce point locations, which are plotted in Figure~\ref{fig:GNSS_Ground_Tracks_CHTI}, and azimuth and elevation data. To approximate Vertical TEC (VTEC), STEC data were multiplied by $\sin \left ( \alpha \right )$, where $\alpha$ was the elevation angle of the satellite measured at the ground station. This quantity will be referred to here as Pseudo-VTEC (PVTEC). Data were recorded at $30$~s intervals, though the time series was interspersed with brief gaps in availability.

\begin{figure}
\includegraphics[width=120mm]{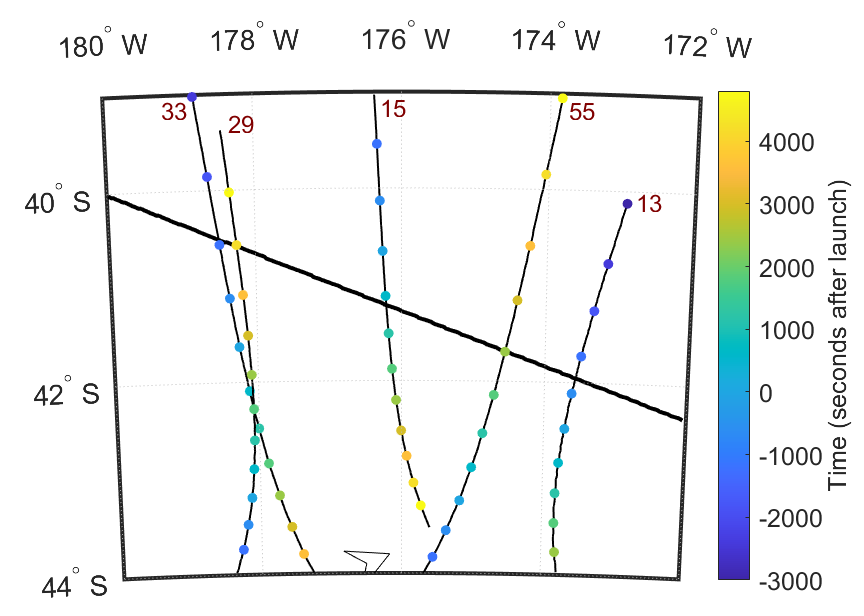}
\centering
\caption{Similar to Figure~\ref{fig:GNSS_Ground_Tracks_CHTI}, zoomed in around pierce point crossings of the ground track.}
\label{fig:GNSS_Ground_Tracks_CHTI_zoom}
\end{figure}

The trajectory of the rocket passed to the north of the islands and is shown in Figure~\ref{fig:GNSS_Ground_Tracks_CHTI}. In the $90$~minutes following launch, satellite 15, satellite 29, and satellite 55 pierce points were identified as crossing the rocket ground track and selected for further study. As shown in Figure~\ref{fig:GNSS_Ground_Tracks_CHTI_zoom}, the crossing occurred earliest for satellite 15, followed by satellite 55, and then satellite 29.

To provide comparison with an undisturbed ionosphere, data at a time offset of $1$~day (i.e. at the same time of day for March 23) were also considered. Levels of solar activity were similarly low for both days, with observed Penticton $F_{10.7}$ solar flux index of $80$~SFU and $79$~SFU on the launch and following day respectively \cite{tapping2013}. There were no flares of Class C or greater on either day. Geomagnetic activity was also low on both days, with the Potsdam $K_p$ index not exceeding $3+$ on either day \cite{matzka2021}. As the orbital period of GNSS satellites is half a sidereal day, the ground track of each GPS satellite approximately repeats from one day to the next. By contrast, the ratio of the orbital period of GLONASS satellites to a sidereal day is approximately $8 / 17$. Because $8$ GLONASS satellites orbit in each plane, a satellite will approximately repeat the ground track traced by the preceding satellite the previous day. Thus, data for satellite 55 on the launch day could be compared with those for satellite 48 the following day.
% GPS satellites repeat ground track with a small drift in local time
% 2.125147134 is the ratio of sidereal days to GLONASS periods

TEC data from the OWMG reference station, located at $44.024^\circ$S, $176.369^\circ$E, $21.620$~m altitude, were also analysed. The features of these data were very similar to those of the nearby CHTI station, supporting the same conclusions but not providing additional information. Therefore, OWMG station data are not presented here.
% OWMG located at $44.024^\circ$S, $176.369^\circ$E, $21.620$~m altitude.

\section{Results}
\label{sec:results}
\begin{figure}
\includegraphics[width=150mm]{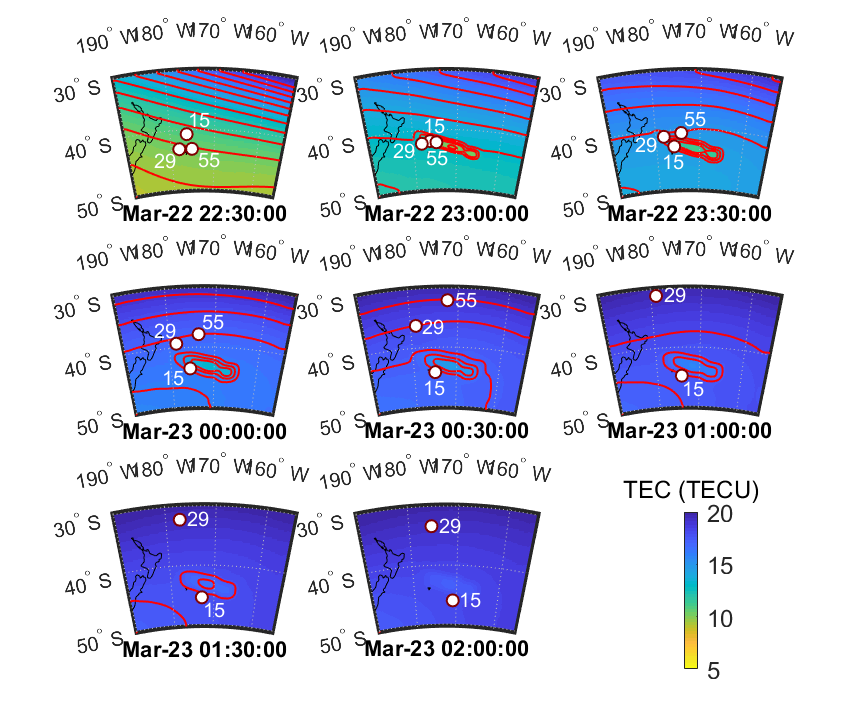}
\centering
\caption{VTEC maps from the GITM simulation of the Electron launch at $22$:$30$~UTC on March 22, 2021. Red contour lines occur at intervals of $1$~TECU~$\equiv 10^{16} \text{electrons.m}^{-2}$. Estimated ionospheric pierce points for satellites 15, 55, and 29 are overlayed.}
\label{fig:TEC_maps}
\end{figure}

The initial expansion and subsequent decay of the simulated ionospheric depletion is shown in Figure~\ref{fig:TEC_maps}. Shortly after launch, the VTEC depletion was highly elongated and was aligned with the rocket ground track. At later times the depletion expanded perpendicular to the ground track while the magnitude at its centre decreased. Comparing simulations with and without rocket exhaust, the largest VTEC depletion was $3.46$~TECU, occurring at $23$:$30$. The background VTEC in the simulation increases and becomes more spatially uniform as local time progresses from late morning to mid afternoon ($11$:$15$ to $14$:$45$ CHAST).
% -3.4564
% 11:15 to 14:45 

\begin{figure}
\includegraphics[width=100mm]{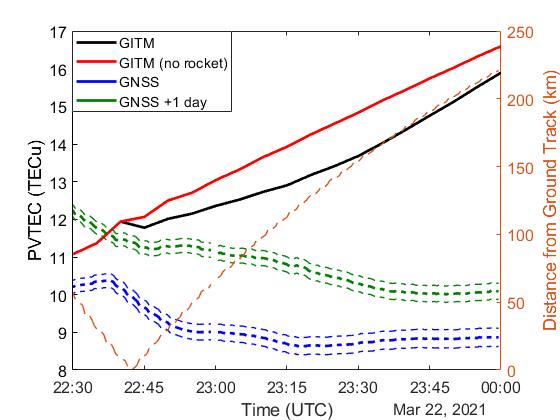}
\centering
\caption{PVTEC measured at the CHTI ground station for GNSS satellite 15. Simulated results from GITM with (black line) and without (red line) the addition of rocket exhaust are shown for a $90$-minute period following the launch. GNSS measurements from the satellite (blue lines) and comparable measurements from the next day (green lines) are also provided. Mean value (thick line) and upper and lower bounds (thin lines) are plotted for the GNSS measurements. Estimated distance from the pierce point to the ground track is also indicated (thin orange line, right axis)}
\label{fig:SATNO15_CHTI_VTEC}
\end{figure}

\begin{figure}
\includegraphics[width=100mm]{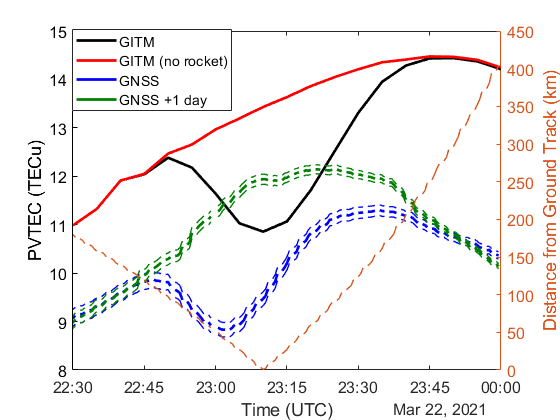}
\centering
\caption{Similar to Figure~\protect\ref{fig:SATNO15_CHTI_VTEC} for GNSS satellite 55 (black, red, and blue lines) and satellite 48 (green lines).}
\label{fig:SATNO55_CHTI_VTEC}
\end{figure}

\begin{figure}
\includegraphics[width=100mm]{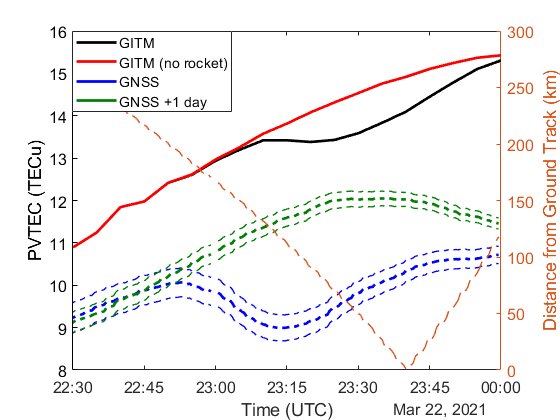}
\centering
\caption{Similar to Figure~\protect\ref{fig:SATNO15_CHTI_VTEC} for GNSS satellite 29.}
\label{fig:SATNO29_CHTI_VTEC}
\end{figure}

Figures~\ref{fig:SATNO15_CHTI_VTEC}, \ref{fig:SATNO55_CHTI_VTEC}, and \ref{fig:SATNO29_CHTI_VTEC} compare the GITM output with GNSS measurements taken at the CHTI ground station. In each case, GITM output is shown with and without rocket exhaust gasses while GITM observations are shown on the day of launch and following day. Evidence of a depletion in $\text{STEC} \times \sin \left ( \alpha \right )$ was found for GNSS data following the launch in each case. The depletion appears earliest for satellite 15, followed by satellite 55, and then satellite 29.
% Similar results were obtained for the nearby OWMG ground station.

% Make TECU capatilisation consistent with text

The estimated pierce point for satellite 15 observations crossed the ground track at approximately $22$:$40$~UTC, heading southward. Therefore, STEC measurements around this time reflected ionosphere changes near the centre of the exhaust plume shortly after passage of the rocket. Figure~\ref{fig:SATNO15_CHTI_VTEC} shows that a sudden variation in the rate of change in PVTEC occurred around $22$:$40$~UTC for both simulated GITM and real GNSS observations. Between $22$:$40$ and $22$:$50$~UTC, the observed quantity changes at approximately $-0.1 \text{ TECU.min}^{-1}$. During the same period, the difference between the simulated quantity with and without exhaust changed at approximately $-0.05 \text{ TECU.min}^{-1}$. The background value of PVTEC steadily increased in the simulation without exhaust, but declined slightly in the real data offset by $1$~day.
% Figure~\ref{fig:SATNO15_CHTI_VTEC} shows how electron densities along the line of sight decreased rapidly, starting shortly after the 
% GITM indicates that the line of site continued to pass through the depletion throughout the observation period

For satellite 55 observations, the pierce point crossed the ground track at approximately $23$:$10$~UTC. Figure~\ref{fig:SATNO55_CHTI_VTEC} shows that this crossing approximately coincided with a minimum in PVTEC for the GITM simulation. However, the real GNSS observations indicate that the minimum occurred earlier, at $23$:$02$~UTC. The simulated maximum depletion was $2.6$~TECU while the observed maximum (based on the difference between the $1$~day offset and post-launch data) was $2.7$~TECU. The depletion in the simulated quantity was apparent ($>1\%$ of maximum value) in the GITM simulations between $22$:$45$ and $23$:$45$~UTC. Significant differences between the real GNSS data and those offset by $1$~day were observed for approximately the same time period.

In the case of satellite 29, the ground track crossing occurred at approximately $23$:$40$~UTC. However, the minimum in PVTEC for the GITM simulation occurred earlier at $23$:$20$~UTC. Observed values of this quantity had a minimum at $23$:$14$~UTC. The maximum depletion based on the simulations was $0.9$~TECU, while the maximum depletion based on observations (determined as above) was $2.6$~TECU. The depletion appeared in simulated data from $23$:$00$~UTC onwards (at $>1\%$ of maximum value) and significant differences between real GNSS data and those offset by $1$~day appeared at approximately this time.

\begin{figure}
\includegraphics[width=100mm]{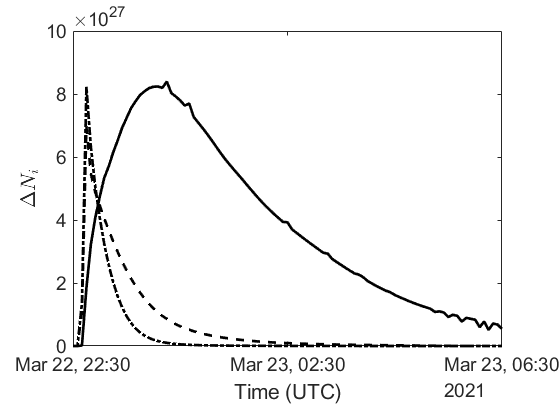}
\centering
\caption{Changes in the total number of electrons and exhaust molecules above $200$~km altitude in the GITM simulation due to the rocket. $\Delta N_{e}$ (solid line), $\Delta N_{\text{CO}_2}$ (dashed line), and $\Delta N_{\text{H}_2\text{O}}$ (dash-dotted line) are shown.}
\label{fig:species_number}
\end{figure}

Global evolution of the depletion is illustrated by plotting changes in the total numbers of exhaust molecules and electrons in Figure~\ref{fig:species_number}. We define $-\Delta N_e$, $\Delta N_{\text{CO}_2}$, and $\Delta N_{\text{H}_2\text{O}}$ as the respective differences in the number of electrons, $\text{CO}_2$ molecules, and $\text{H}_2\text{O}$ molecules due to the addition of rocket exhaust. Only electrons and molecules above $200$~km are considered, as significant ionospheric depletions are limited to this region and GITM introduces unrealistic changes in concentrations near the lower boundary at $100$~km \cite{bowden2020}. $\Delta N_{\text{CO}_2}$, and $\Delta N_{\text{H}_2\text{O}}$ both decreased rapidly from their initial value after deposition by the rocket. This resulted in a maximum in $\Delta N_e$ of approximately $8.2 \times 10^{27}$ at $00$:$05$~UTC on March 23. Subsequently, $\Delta N_e$ decayed with a characteristic time of $\tau \approx 2$ to $3$~hours ($\frac{1}{\tau} = \frac{1}{\delta n} \frac{d \Delta n}{\delta t}$).
% (here is defined as the difference between the number of electrons in the simulation without and with rocket exhaust).
% Due to a combination of chemical reactions, gravity, and diffusion
% tau

\begin{figure}
\includegraphics[width=80mm]{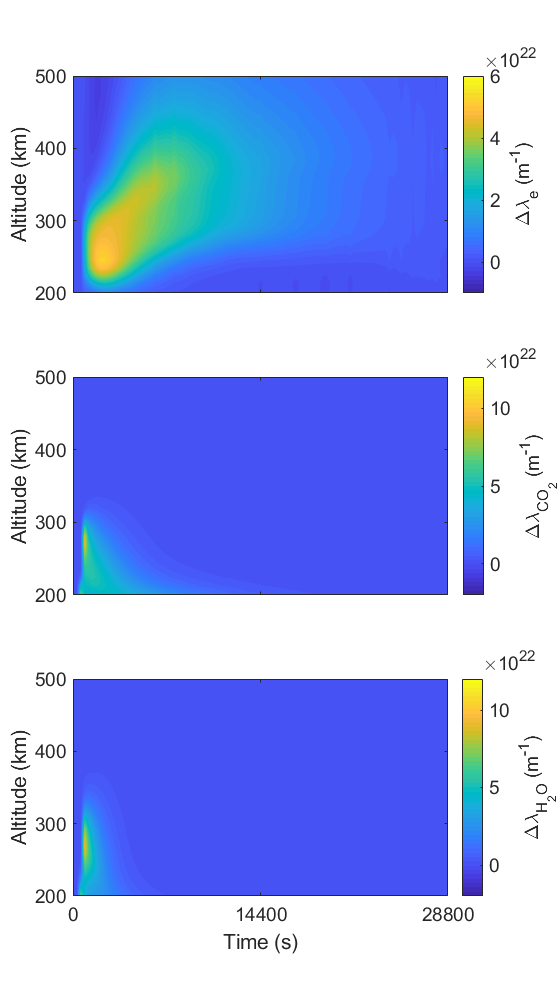}
\centering
\caption{Altitude distribution of changes in the number of (a) electrons, (b) $\text{CO}_2$ molecules, and (c) $\text{H}_2\text{O}$ molecules over time.}
\label{fig:altitude_density}
\end{figure}

Simulated changes in the number of exhaust molecules and electrons across different altitudes resulting from addition of rocket exhaust are shown in Figure~\ref{fig:altitude_density}. We define $-\lambda_e$, $\lambda_{\text{CO}_2}$, and $\lambda_{\text{H}_2\text{O}}$ as the respective differences in the number of electrons, $\text{CO}_2$ molecules, and $\text{H}_2\text{O}$ molecules per unit altitude due to the addition of rocket exhaust. Both $\text{CO}_2$ and $\text{H}_2\text{O}$ concentration increases were initially peaked around $270$~km altitude before falling to lower altitudes. Electron concentration decreases were initially concentrated around similar altitudes, before they diffused to higher altitudes over subsequent hours. These decreases tended to persist for longer times at higher altitudes.

% Include a) b) and c) labels in figure

\section{Discussion}
\label{sec:discussion}
% Depletion decays approximately exponentially after loss of exhaust species (check)
The observations and simulations presented in Section~\ref{sec:results} showed that ionospheric effects of even comparatively small launch vehicles such as the Electron are detectable through GNSS observations. Both real and simulated measurements evidinced ionospheric depletions larger than the uncertainty for these measurements. These results suggest that observations from GNSS ground stations are sufficiently sensitive to detect most space launch vehicles as they pass through F-region altitudes. Numerical simulations could be used in future to provide data for a classification scheme to determine when launches occur using GNSS observations.

% More consistent with observations than example studied previously
% Horizontal trajectory means that advection to different altitudes is less important than cases previously examined
% Comparatively short-lived exhaust species
Our method provides better agreement with observations of the magnitude of TEC depletion in the current case than it did for the Falcon 9 launch of FORMOSAT-5 which was investigated previously \cite{bowden2020}. This can be attributed to the shallower angle of ascent for the case presented in this paper. Consequently, the vertical component of initial plume velocity was lower in the Electron case and therefore ignoring the consequent vertical advection of $\text{CO}_2$ and $\text{H}_2\text{O}$ molecules in the simulations is more realistic. Lower altitudes corresponded to decreased thermosphere residence times for the molecules. Moreover, ionosphere production rates increase with decreasing altitude in the F-region. Together these effects result in a shorter-lived ionospheric depletion (comparing Figure~\ref{fig:species_number} with Figure 10(b) in \citeA{bowden2020}). In future, satellite-based GNSS radio occultation may provide opportunities to study the evolution in the depletion at higher altitudes more directly.
% Could reference Richards and Torr 1988 for ion production rates

Advection of rocket exhaust due to its initial velocity in the horizontal direction may have been important in the Electron case. This may help explain why the depletion appeared to have been underestimated by GITM for the satellite 29 observations but not those for satellite 55. Figure~\ref{fig:GNSS_Ground_Tracks_CHTI} shows that the pierce point for the former satellite passed closer to the launch site. The Electron $I_{sp}$ value corresponded to an exhaust velocity of $v_e = 3360 \textup{ m.s}^{-1}$. The launch vehicle was estimated to reach this speed approximately $297$~s after launch when located at $40.88^\circ$S, $177.31^\circ$W. Prior to this time, exhaust would have had an initial velocity opposite that of the launch vehicle (in an Earth Centred Earth-fixed reference frame).
% Consider estimated position when rocket and exhaust velocity are equal

% Smaller rocket plume exerts less force on the background atmosphere
% Approximation of point release 
% Less opportunity for diffusion prior to chemical reaction
The launch of an Electron SLLV better approximates our model of initial plume expansion as a series of point releases than those of larger launch vehicles. Plume dimensions are reduced for smaller launch vehicles, with an approximate analytical treatment indicating length and maximum radius are each proportional to $\sqrt{T}$ where $T$ is thrust \cite{jarvinen1966}. Moreover, forces exerted by smaller launch vehicles on the background ionosphere and thermosphere are expected to affect the background state less. Resulting disturbances such as the "snow-plow effect", wherein the plume directly displaces the ionosphere, are thus minimised.

For smaller launch vehicles initial high concentrations of exhaust gasses are less likely to saturate the ionosphere, reducing electron and ion concentrations to very low levels thereby inhibiting these gasses further contribution to the depletion. Therefore, in such cases, numerical models will be less sensitive to the treatment of the early expansion of the plume. It will also limit the opportunity exhaust gasses have to diffuse to higher altitudes where they can give rise to long-lived ionospheric depletions. Figure~\ref{fig:species_number} shows that time that exhaust gasses spent within the thermosphere in the simulation was short compared with that for which the depletion existed ($\Delta N_{\textup{CO}_2}$ and $\Delta N_{\textup{H}_2\textup{O}}$ declined below $10\%$ of peak levels $95$~min and $65$~min after launch respectively) for the Electron launch. The exhaust gasses declined more rapidly in simulations for this case than for either Falcon 9 launch examined by \citeA{bowden2020} (see Figure~10 in the reference).

The minima in PVTEC observations of satellites 55 and 29 occurred earlier than their pierce points crossed the satellite track. Therefore, the depletion appears to have been pushed southwards towards the CHTI ground station due to advection of rocket exhaust gasses by thermospheric winds. GITM estimated that minima occurred later, suggesting the model underestimated the southward component of these winds.
% The largest decreases in $\textup{STEC} \times \sin \left ( \alpha \right )$ inferred from observations of satellites 55 and 29 occurred earlier than those calculated through simulations.
% As the pierce points for observations from both satellites moved approximately northward, the depletion appears to have been advected southwards.

% This seems to gradually smooth out gradients compared with those in the initial IRI estimate
% Global convection is not satisfactorily modelled for the regional simulation
The regional GITM simulations in this study did not include global-scale convection, which affects neutral wind and ion drift velocities. Open boundary conditions at the upper and lower bounds for latitude and longitude were applied in these simulations, which produces unrealistic results for longer simulations. Figure~\ref{fig:TEC_maps} shows that latitudinal variation in TEC outside of the depletion decreases significantly over time. The boundary condition issue could be rectified in future by using global GITM runs with local grid refinement, a capability for investigating multi-scale phenomena described by \citeA{zhao2020}.
% Deng 2018 claim 6 hours is maximum realistic time, reported by Zhao
% IRI implies the gradient should not change significantly over this time

\section{Conclusions}
\label{sec:conclusions}
Ionospheric depletions were observed following the launch of an Electron SLLV. Observations and GITM simulations of STEC changes due to these depletions were comparable in magnitude and duration, demonstrating the promise of GCMs for modelling these anthropogenic impacts upon the ionosphere. Simulations indicated that exhaust gasses were short-lived compared with the resulting ionospheric depletion, unlike the FORMOSAT-5 launch case previously examined by \citeA{bowden2020}. Evidence was found in GNSS observations for southward advection of the ionospheric depletion, which was not accounted for in the regional GITM simulation. These findings will inform future GCM development for simulating interactions between rockets and the ionosphere.

\acknowledgments
This work was carried out with funding from the Air Force Office of Scientific Research (award number FA2386-22-1-4003). Resources from the National Computational Infrastructure (NCI), supported by the Australian Government, were used. MAPGPS TEC data were obtained through the Madrigal CEDAR Database,accessible via the World Wide Web (at \url{http://cedar.openmadrigal.org/index.html/}).

%% ------------------------------------------------------------------------ %%
%% References and Citations

\bibliography{ionospheric_hole_paper}

%Reference citation instructions and examples:
%
% Please use ONLY \cite and \citeA for reference citations.
% \cite for parenthetical references
% ...as shown in recent studies (Simpson et al., 2019)
% \citeA for in-text citations
% ...Simpson et al. (2019) have shown...
%
%
%...as shown by \citeA{jskilby}.
%...as shown by \citeA{lewin76}, \citeA{carson86}, \citeA{bartoldy02}, and \citeA{rinaldi03}.
%...has been shown \cite{jskilbye}.
%...has been shown \cite{lewin76,carson86,bartoldy02,rinaldi03}.
%... \cite <i.e.>[]{lewin76,carson86,bartoldy02,rinaldi03}.
%...has been shown by \cite <e.g.,>[and others]{lewin76}.
%
% apacite uses < > for prenotes and [ ] for postnotes
% DO NOT use other cite commands (e.g., \citet, \citep, \citeyear, \citealp, etc.).
% \nocite is okay to use to add references from your Supporting Information
%

\end{document}